\documentclass[prb,superscriptaddress,twocolumn,floatfix]{revtex4-2}
\usepackage[utf8]{inputenc}
\usepackage[T1]{fontenc}
\usepackage{mathptmx}
\usepackage{rotating}
\usepackage{listings}
\usepackage{xcolor}
\usepackage{siunitx}
\usepackage{lipsum}
\usepackage{wrapfig}
\usepackage{pgfplots}
\pgfplotsset{compat=1.17}
\usepgfplotslibrary{external}
\pgfplotsset{plot coordinates/math parser=false}
\usetikzlibrary{plotmarks}
\usepgfplotslibrary{colormaps}
\definecolor{codegreen}{rgb}{0,0.6,0}
\definecolor{codegray}{rgb}{0.5,0.5,0.5}
\definecolor{codepurple}{rgb}{0.58,0,0.82}
\definecolor{backcolour}{rgb}{0.95,0.95,0.92}
\lstdefinestyle{mystyle}{
    backgroundcolor=\color{backcolour},   
    commentstyle=\color{codegreen},
    keywordstyle=\color{magenta},
    numberstyle=\tiny\color{codegray},
    stringstyle=\color{codepurple},
    basicstyle=\ttfamily\footnotesize,
    breakatwhitespace=false,         
    breaklines=true,                 
    captionpos=b,                    
    keepspaces=true,                 
    numbersep=5pt,                  
    showspaces=false,                
    showstringspaces=false,
    showtabs=false,                  
    tabsize=2
}

\usepackage{times,amsmath}
\usepackage{epsfig}
\usepackage{color}
\usepackage{ulem}
\usepackage{graphicx}% Include figure files\
\usepackage{dcolumn}% Align table columns on decimal point
\usepackage{bm}% bold math
\usepackage{bookmark}
\usepackage{tabularx}% bold math
\usepackage{hyperref}
\usepackage{multirow}
\usepackage{booktabs}
\usepackage{pgfplots}
\usepackage{pgfplotstable}
\usepackage{tikz}
\usetikzlibrary{shapes.geometric,positioning,arrows.meta}
\hypersetup{colorlinks=true, citecolor=blue, filecolor=blue, linkcolor=blue, urlcolor=blue}

\usepackage[switch, pagewise]{lineno}
\linenumbers\relax 
\nolinenumbers

\begin{document}

\title{Accelerated Inorganic Electrides Discovery by Generative Models and Hierarchical Screening}

\author{Shuo Tao}
\email{stao@charlotte.edu}
\affiliation{Department of Mechanical Engineering and Engineering Science, University of North Carolina at Charlotte, Charlotte, NC 28223, USA}

\author{Qiang Zhu}
\email{qzhu8@charlotte.edu}
\affiliation{Department of Mechanical Engineering and Engineering Science, University of North Carolina at Charlotte, Charlotte, NC 28223, USA}
\affiliation{North Carolina Battery Complexity, Autonomous Vehicle and Electrification (BATT CAVE) Research Center,
Charlotte, NC 28223, USA}

\date{\today}

\begin{abstract}
Electrides are exotic compounds in which excess electrons occupy interstitial regions of the crystal lattice and serve as anions, exhibiting exceptional properties such as low work function, high electron mobility, and strong catalytic activity. Although they show promise for diverse applications, identifying new electrides remains challenging due to the difficulty of achieving energetically favorable electron localization in crystal cavities. Here, we present an accelerated materials discovery framework that combines physical principles, diffusion-based materials generation with hierarchical thermodynamic and electronic structure screening. Using this workflow, we systematically explored 1,510 binary and 6,654 ternary chemical compositions containing excess valence electrons from electropositive alkaline, alkaline-earth, and early transition metals, and then filtered them with a high throughput validation on both thermodynamical stability and electronic structure analysis. As a result, we have identified 264 new electron rich compounds within 0.05~eV/atom above the convex hull at the density functional theory (DFT) level, including 13 thermodynamically stable electrides. Our approach demonstrates a generalizable strategy for targeted materials discovery in a vast chemical space.
\end{abstract}

\maketitle

\section{Introduction}
Electrides represent a unique class of ionic crystals in which electrons occupy interstitial regions of the crystal lattice and serve as the nucleus-free anionic species~\cite{dyeElectridesIonicSalts1990,dyeElectridesEarlyExamples2009}. Since the first crystalline electride Cs$^+$(18-crown-6)$_2$e$^-$ was reported~\cite{dawesFirstElectrideCrystal1986}, these materials have attracted significant attention owing to their unusual electronic properties and promising applications in various electronic devices and chemical reactions~\cite{hosonoAdvancesMaterialsApplications2021}.

While early electride research focused primarily on organic compounds~\cite{huangStructureK+cryptand222JElectride1988,singhTheoreticalDeterminationThat1993,lePreparationAlkalideElectride1982}, these materials suffer from thermal instability above room temperature and sensitivity to moisture, severely limiting their practical applications~\cite{dyeElectridesEarlyExamples2009}. In 2003, Hosono and coworkers~\ achieved
a significant progress toward the realization of inorganic electrides was achieved by the successful synthesis of Ca$_6$Al$_7$O$_{16}$ (C12A7:2e$^-$) via oxygen-reducing processes from the mineral mayenite (12CaO$\cdot$7Al$_2$O$_3$)~\cite{Matsuishi-Science-2003}. The discovery of thermally stable C12A7:2e$^-$ has stimulated extensive efforts to exploit these materials to enable new technology applications such as ammonia synthesis and ~\cite{Kitano-NChem-2012, Kanbara-JACS-2015, Kitano-NC-2015} and electron emission \cite{drobny2024endurance}, as well as to search for other inorganic electrides~\cite{hosonoAdvancesMaterialsApplications2021, Li-Review-2016, liu2020electrides}.

Based on the dimensionality of anionic electron distribution, electrides can be classified into zero-dimensional (0D), one-dimensional (1D), two-dimensional (2D), and three-dimensional (3D) categories, each exhibiting unique electronic and transport characteristics.
Among these, the 2D electride Ca$_2$N has attracted considerable attention due to its high electron mobility, low work function, and striking anisotropic magnetoresistance~\cite{leeDicalciumNitrideTwodimensional2013,druffelExperimentalDemonstrationElectride2016}.
The Ca$_2$N structure type ($R\overline{3}m$: 166) with an anti-CdCl$_2$ layered structure represents a particularly promising structural family for 2D electride formation, including Sr$_2$N, Ba$_2$N, and Y$_2$C~\cite{tadaHighThroughputInitioScreening2014, Ming-JACS-2016, zhangComputerAssistedInverseDesign2017}.
Similarly, the 1D electrides of the Mn$_5$Si$_3$-type have been widely studied due to their peculiar semiconducting and superconducting behaviors~\cite{wangExplorationStableStrontium2017, Lu-JACS-2016, zhangElectrideSuperconductivityBehaviors2017, zhangElectronConfinementChannel2015}.

Beyond the chemical novelty, electrides have also emerged as a promising platform for realizing topological electronic phases. Recent theoretical work has demonstrated that the weakly bound anionic electrons near the Fermi level in electrides naturally facilitate band inversions necessary for topological insulating and semimetal phases. Several electrides including Y$_2$C, Sc$_2$C, Sr$_2$Bi, and HfBr have been predicted to exhibit nontrivial topological properties such as nodal-line semimetal states, quantum spin Hall effects, and even quantum anomalous Hall insulator behavior~\cite{hirayamaElectridesNewPlatform2018, Park-PRL-2018,
huangTopologicalElectrideY2C2018}. High-pressure studies have further expanded the electride landscape, revealing that simple metals like sodium and lithium can transform into insulating electride phases under extreme compression, where valence electrons migrate to interstitial sites and form quasi-atomic or quasi-molecular units~\cite{maTransparentDenseSodium2009,miaoHighPressureElectrides2014,miaoHighPressureElectridesChemical2015,wanHighpressureElectridesDesign2019,naumovMetallicSurfaceStates2017}.

To date, the number of experimentally confirmed electrides remains small compared to the vast chemical space of potential candidates. This scarcity has motivated further computational efforts for new electride discovery through high-throughput screening of existing crystal structure databases~\cite{tadaHighThroughputInitioScreening2014,burtonHighThroughputIdentificationElectrides2018,zhouDiscoveryHiddenClasses2019,zhuComputationalDiscoveryInorganic2019}.
Tada et al.~pioneered high-throughput screening of approximately 34,000 materials from the Materials Project, employing three key indicators: positive total formal charge, layered structures for two-dimensionality, and empty interlayer spaces to search for new 2D electrides~\cite{tadaHighThroughputInitioScreening2014}. 
Building on this foundation, Burton et al.~performed automated density functional theory (DFT) calculations on over 60,000 compounds from the Materials Project to identify 65 new electride candidates~\cite{burtonHighThroughputIdentificationElectrides2018}. 
This number was further expanded to 167 by Zhu et al. through a more comprehensive pipeline that accounts for electron localization analysis in both spatial and energy space~\cite{zhuComputationalDiscoveryInorganic2019, wang2018ternary}. Additionally, structure engineering based on existing materials has been applied to discover new electrides using simple structural prototypes~\cite{inoshitaExplorationTwoDimensionalElectrides2014, mcrae2022sc2c} or complex zeolite frameworks~\cite{kang2023first,liu2025semiconducting}.

While database screening have proven invaluable for identifying electrides in existing material prototypes, it is also appealing to design and discover new electrides in unexplored chemical space. Hence, crystal structure prediction methods based on global optimization algorithms have been attempted in various ways~\cite{zhangComputerAssistedInverseDesign2017, wangExplorationStableStrontium2017,zhuComputationalDesignFlexible2019, Ming-JACS-2016}. Zhang et al.~developed an inverse-design method to identify 24 stable and 65 metastable new inorganic electrides covering 33 distinct prototypes~\cite{zhangComputerAssistedInverseDesign2017}. Wang et al.~combined ab initio structure searches with experimental validation to synthesize Sr$_5$P$_3$, an 1D semiconducting electride~\cite{wangExplorationStableStrontium2017}. Zhu et al.~computationally designed flexible electrides Rb$_3$O and K$_3$O with nontrivial band topology, demonstrating that multiple types of cavities enable tuning of anionic electron spatial arrangements~\cite{zhuComputationalDesignFlexible2019}. These results collectively demonstrate that it is possible to identify truly new electrides that may not been synthesized in the past. However, they were mostly limited to a few chemical systems and compositions due to the concerns of computational cost.

Despite these exciting developments in theory-driven discovery~\cite{wangExplorationStableStrontium2017, chanhom2019sr3crn3, mcrae2022sc2c}, identifying new electrides remains a grand challenge due to three main obstacles. First, determining which chemical space to explore is critical since it is infeasible to systematically sample all possible combinations across the whole periodic table of over 100 elements. Once the chemical space is determined, it is essential to develop a sustainable framework capable of rapidly performing two critical tasks: (1) predicting all energetically favorable crystal structures within a given chemical space, and (2) evaluating their thermodynamic feasibility and relevant properties. To date, most existing methods applied to electride research heavily rely on DFT-based electronic structure calculations for both tasks. But they are not suitable for systematic exploration study due to their prohibitive computational costs.

Fortunately, recent advances in machine learning (ML) and artificial intelligence (AI) have revolutionized materials discovery and evaluation~\cite{merchant2023scaling, zeni2025generative}. Diffusion-based generative models~\cite{DiffCSP, zeni2025generative, levy2024symmcd, miller2024flowmm} enable efficient exploration of crystal structure space by learning data-driven representations, generating diverse low-energy structures orders of magnitude faster than traditional global optimization methods. Concurrently, ML foundation model interatomic potentials~\cite{yang2024mattersim, batatia2025foundation, wood2025family} have dramatically accelerated structural evaluation, providing near-DFT accuracy at computational costs reduced by three to four orders of magnitude. This convergence of advances in generative AI and ML potentials enables targeted exploration of well-defined chemical spaces that would be computationally prohibitive using conventional approaches. While some concerns have been raised regarding the novelty of AI-based crystal generation~\cite{cheetham2024artificial}, these methods have been applied to catalyze discoveries across diverse functional materials including magnets~\cite{zeni2025generative} and thermoelectrics~\cite{ghafarollahi2025autonomous, guo2025generative, longInverseDesignHighperformance2025}.

%: diffusion-based generative models for crystal structure prediction and machine learning interatomic potentials for rapid thermodynamic screening. Generative models trained on large crystallographic databases can produce diverse, chemically reasonable structures for specified compositions without requiring computationally expensive iterative optimization, fundamentally changing the economics of candidate structure generation. When coupled with machine learning potentials that provide near-DFT accuracy at computational costs reduced by three to four orders of magnitude, a hierarchical screening workflow becomes feasible where thousands of generated structures can be rapidly evaluated for thermodynamic stability before committing expensive DFT resources to only the most promising candidates. This prescreening approach addresses the computational bottleneck that previously limited electride discovery efforts to either database mining or exhaustive DFT calculations on relatively small composition sets. However, successful implementation requires careful validation that the machine learning prescreening maintains sufficient accuracy to avoid both false negatives (discarding stable electrides) and false positives (promoting unstable structures to expensive DFT validation), necessitating systematic benchmarking against ground-truth DFT results.

In this work, we demonstrate a new computational framework for efficient exploration of large chemical space, using inorganic electride discovery as a case study. Crucially, we began with chemical space selection based on physical principles of electride formation, focusing on binary and ternary compositions with excess valence electrons from electropositive metals. We then employed \texttt{MatterGen} generative model~\cite{zeni2025generative} and foundation interatomic potentials \texttt{MatterSim}~\cite{yang2024mattersim} to rapidly screen structures in the reduced chemical space. This was further integrated with our previously developed high-throughput electronic structure framework~\cite{zhuComputationalDesignFlexible2019} to identify low-energy electrides. Our results demonstrate the power of generative AI and ML to accelerate targeted materials discovery, provide a generalizable strategy 
for future explorations beyond electrides.

The remainder of this paper is organized as follows. Section~\ref{sec:methods} describes our computational methodology in detail, including targeted chemical space selection, generative structure prediction, and the hierarchical screening workflow combining ML prescreening with DFT validation for electride identification. Section~\ref{sec:results} presents our discoveries of binary and ternary electride candidates, analyzing their structural characteristics, thermodynamic stability, and electronic structure features, with emphasis on interstitial electron localization patterns and prototype recurrence across chemical space. Section~\ref{sec:conclusion} discusses the implications of our findings for rational electride design principles, followed by concluding remarks and perspectives on how generative models and machine learning potentials enable targeted exploration of vast chemical space for accelerated materials discovery.

\section{Computational Methods}\label{sec:methods}

\begin{figure*}[t]
\centering
\includegraphics[width=0.9\textwidth]{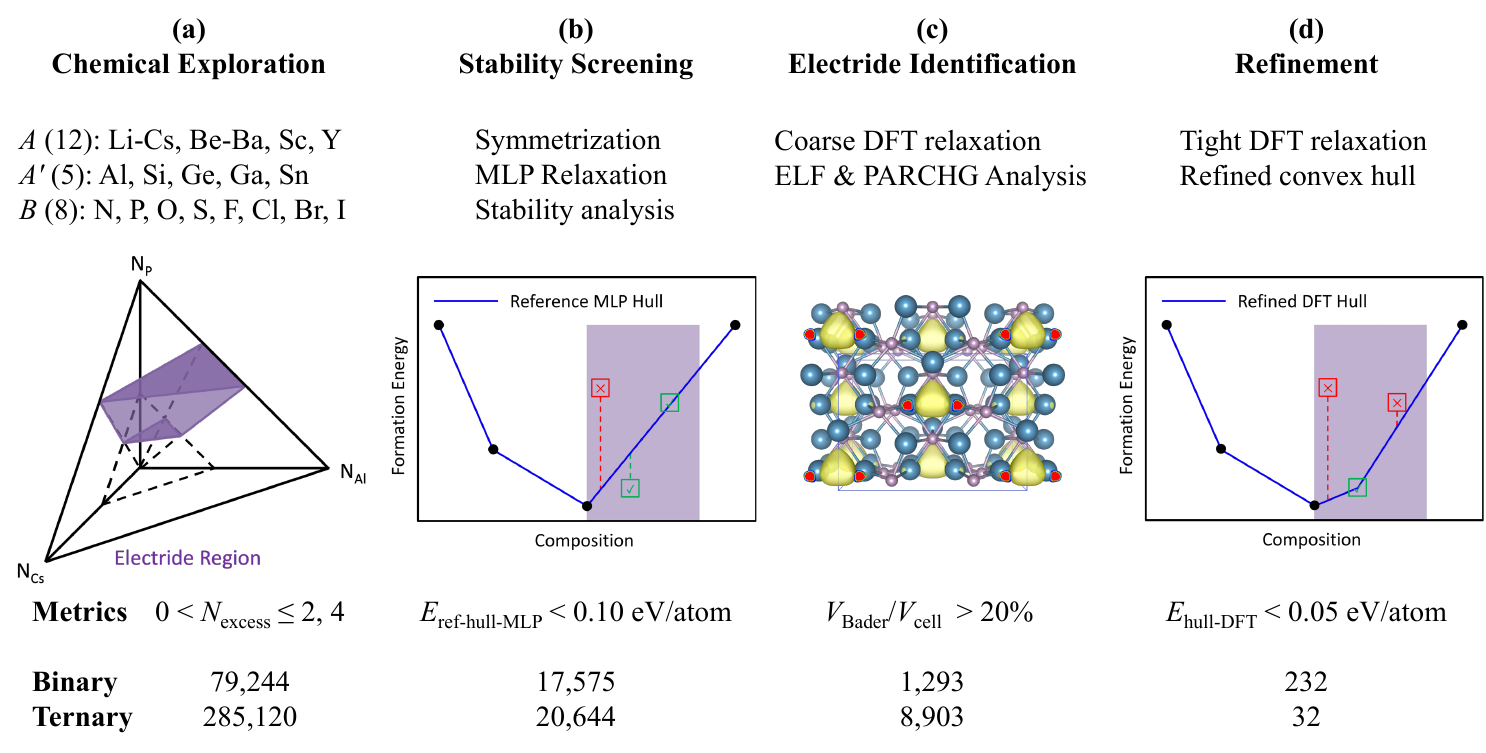}
\vspace{-3mm}
\caption{The workflow for accelerated electride discovery combining physical principles, generative modeling (\texttt{MatterGen}), machine learning potential prescreening (\texttt{MatterSim}), and high-throughput DFT validation for electride and stability analysis. The top panel outlines key operations at each stage; the middle panel illustrates the definition of metrics used for candidate selection with the targeted compositions being highlighted by the light purple color; and the bottom panel specifies filtering criteria and the number of candidates retained after each stage.}
\label{fig:workflow}
\vspace{-3mm}
\end{figure*}

Fig.~\ref{fig:workflow} outlines our electride discovery workflow, including: (1) generating candidate structures across targeted electron-rich compositions using the \texttt{\texttt{MatterGen}} model; (2) relaxing the generated structures and performing thermodynamical stability analysis to select low energy candidates ($E_\text{ref-hull-MLP}$) via machine learning potentials; (3) performing high-throughput structural optimization and electronic structure analysis using DFT to identify the potential electrides; and (4) conducting high-precision DFT calculations for promising electride candidates to obtain accurate final energetics ($E_\text{hull-DFT}$). Below, we describe each stage in detail.

\subsection*{Stage 1: Chemical Space Exploration}
In the past, crystalline interstitial electrons are commonly found in environments dominated by electropositive metals~\cite{wang2018ternary, zhuComputationalDiscoveryInorganic2019, zhangComputerAssistedInverseDesign2017, dong2022electronegativity}, including alkaline, alkaline-earth, and early transition metals. This observation strongly suggests that these electropositive metals are essential for providing excess electrons and achieving the interstitial anionic electrons in the surrounding low-affinity nucleus. Consequently, we focused our exploration on binary (AB) and ternary (AA$^\prime$B) compounds with excess valence electrons from 12\textit{ A-type} electropositive metals (Li, Na, K, Rb, Cs, Be, Mg, Ca, Sr, Ba, Sc, Y) combined with 8 \textit{B-type} electronegative non-metals (N, P, O, S, F, Cl, Br, I). For ternary compositions, we additionally incorporated 5 \textit{A$^\prime$-type} Earth-abundant elements (Al, Si, Ge, Ga, Sn) to enhance structural and chemical diversity, encompassing a total of 96 binary and 480 ternary chemical systems. Exhaustive enumeration of all possible combinations with up to 20 atoms per unit cell would require sampling 190 discrete $A_mB_n$ ($m+n\leq20$) combinations per binary system and 1,140 $A_mA^{\prime}_lB_n$ ($l+m+n\leq20$) combinations per ternary system, resulting in an astronomically large search space. 

However, one does not have to explore all possible compositions. A common characteristic of electrides is that the number of excess electrons should be not only positive but remain moderate to avoid purely metallic behavior. Consequently, we restrict our search to compositions with no more than 4 excess electrons per formula unit for binaries, and 2 for ternaries. Using empirical valence states (e.g., +2 for Ca and -2 for O), the number of excess electrons per formula unit is calculated as:
$N_{\text{excess}}(A_mB_n) = m V_A + n V_B$ and $N_{\text{excess}}(A_mA^\prime_lB_n) = m V_A + l V_{A^\prime} + n V_B$, where $V_A$, $V_{A^\prime}$, and $V_B$ are the valence states of elements A, A$^\prime$, and B, respectively.
Here all compositions are enumerated as reduced (irreducible) formulas, and $N_{\text{excess}}$ is evaluated per reduced formula unit. Thus a composition such as A$_8$B$_2$ reduces to A$_4$B and is counted only once; its non-reduced multiples enter only later, as supercells, during the generative step. The choice of a tighter window for ternaries ($0 < N_{\text{excess}} \leq 2$) reflects both the much larger ternary search space (1,140 versus 190 stoichiometric combinations per system).
%And the physical observation, confirmed in our Results, that the relevant electride chemistry is concentrated at $N_{\text{excess}} = 1$ (half-filled interstitial band) and $N_{\text{excess}} = 2$ (intrinsically semiconducting electrides); the slightly wider binary window ($\leq 4$) was affordable given the smaller binary space.
This restriction dramatically reduces the combinatorial search space. For example, Fig. \ref{fig:workflow}a schematically shows that applying the constraint $0 < N_{\text{excess}} \leq 2$ to a ternary system (Cs-Al-P) can reduce the number of required stoichiometric combinations from 1,140 to 50. Using this criterion, we kept 1,510 binary and 6,654 ternary compositions for further exploration.

Next, we retrained \texttt{MatterGen} \cite{zeni2025generative}, the state-of-the-art diffusion-based generative model using the Materials Project mp\_20 dataset to generate new crystal candidates. The model was trained from scratch in the crystal structure prediction (CSP) configuration, in which the only conditioning information is the target chemical composition. Importantly, no electride-specific or any other property labels (e.g., work function, band gap, or interstitial-electron descriptors) were used during either training or sampling; the targeting of electrides is achieved solely through the physically motivated composition selection above and the downstream screening below, not by biasing the generative model.
%Training was performed with the official \texttt{mattergen-train} interface using distributed data-parallel optimization on two NVIDIA A100 GPUs with full \texttt{fp32} precision, gradient accumulation over four batches, and a maximum of 1,000 epochs ($\sim$53,000 optimization steps); the checkpoint with the lowest validation loss was used for generation.
For each target composition, we generated structures using multiple formula unit up to the ceiling of 20 atoms per unit cell. Specifically, each reduced formula was expanded to all valid supercell multiples within the 20-atom limit (e.g., a 6-atom $A_4B_2$ cell was generated at 6, 12, and 18~atoms/cell; a 9-atom $A_6B_3$ cell at 9 and 18~atoms/cell; and a 19-atom $A_{10}B_9$ cell at 19~atoms/cell only). The number of structures per composition is scaled with system size according to $N_\text{attempts} = 2 \times N_{\text{atoms per cell}}$ summed over its supercells. This setting eventually leads to a total of 79,244 binary and 285,120 ternary structures for \texttt{MatterGen} generation. Using the NVIDIA A100 GPU, it approximately takes 1,000 GPU hours, which is roughly consistent with the speed of 500 structures per GPU hour as reported in the original paper \cite{zeni2025generative}.

\subsection*{Stage 2: Stability Screening}

It is important to note that \texttt{MatterGen} only generates structures in the primitive cell setting without crystallographic symmetry information. To reduce the load for the followup relaxation, we symmetrized all generated structures using \texttt{PyXtal}~\cite{pyxtal} with a tolerance of 0.05~\AA, then performed structural relaxations with the \texttt{MatterSim} machine learning potential trained on Materials Project data~\cite{yang2024mattersim}. Structures were relaxed in batches of 32 (configurable based on GPU memory availability) with calculator instance reuse to maximize GPU efficiency and minimize memory overhead. The relaxed structures were evaluated against reference convex hulls ($E_\text{ref-hull-MLP}$) using Materials Project data via \texttt{pymatgen}'s Materials API~\cite{pymatgen-2013}. As illustrated in Fig. \ref{fig:workflow}b, a thermodynamically stable compound exhibits $E_\text{ref-hull-MLP} \leq 0$~eV/atom, while structures with $E_\text{ref-hull-MLP} \geq 0.10$~eV/atom can be discarded due to the unfavorable energy upon the decomposition to the reference stable compounds. 

Next, a similarity check using \texttt{pymatgen}'s structure matching algorithm was applied to remove duplicate structures with a site distance of 0.2, a fractional length tolerance of 0.2, and a lattice angle tolerance of $5^\circ$. This stage required approximately 400 GPU hours and yielded 17,575 unique binary and 20,644 ternary structures with $E_\text{ref-hull-MLP} < 0.10$~eV/atom for the next validation stage.

\subsection*{Stage 3: Electride Characterization}

Structures with $E_\text{ref-hull-MLP} < 0.10$~eV/atom were selected for further relaxation using \texttt{VASP} 6.2.1~\cite{Vasp-PRB-1996}, employing the GGA-PBE functional~\cite{PBE-PRL-1996} with PAW pseudopotentials~\cite{PAW-PRB-1994}. This DFT relaxation serves two purposes: first, to verify structural stability and ionic convergence in case the MLP relaxation was insufficient, and second, to re-evaluate the energy above hull using the relaxed structures.
To reduce computational costs, we intentionally employed coarse settings for this stage (PREC = Accurate, EDIFF = 10$^{-6}$~eV, EDIFFG = -0.01~eV/\AA~for convergences, and NSW = 30, ISIF = 3, POTIM = 0.3 for ion relaxation).
This coarse DFT relaxation provides an efficient means to further filter out unstable structures exhibiting either nonphysical geometry or unfavorable energy ranking due to artifacts from the MLP prescreening.

Only successfully relaxed structures were retained for subsequent analysis, following our previously reported methodology for electride identification~\cite{zhuComputationalDiscoveryInorganic2019}. Specifically, we first computed the electron localization function (ELF) to identify whether electrons localize in interstitial sites. Upon identification of interstitial ELF maxima, we proceeded to compute five partial charge density (PARCHG) distributions, grouped into two complementary families: three energy-window files that integrate occupied states within $[E_F-\Delta E, E_F]$ for $\Delta E = 0.025$~eV ($e_{0.025}$), 0.5~eV ($e_{0.5}$), and 1.0~eV ($e_{1.0}$), which probe how the interstitial density builds up over progressively deeper energy ranges below the Fermi level; and two band-decomposed files for the top two valence bands ($band_0$ for the VBM and $band_1$ for VBM$-$1), which isolate the band-resolved character of the interstitial electrons. For each PARCHG, the Bader algorithm partitioned electrons into basins based on zero-flux surfaces in the electron density gradient using the modified Bader utility~\cite{Henkelman-CMS-2006}, and the volume of any non-nuclear (interstitial) basin was recorded for each of the five files. A structure was first identified as an electride candidate when an interstitial basin was detected (non-zero volume) in at least one of the two near-Fermi energy windows and in either the deepest energy window or the top valence band, i.e., $(e_{0.025} > 0$ or $e_{0.5} > 0)$ and $(e_{1.0} > 0$ or $band_0 > 0)$. Requiring consistency between the energy-resolved and band-resolved descriptions ensures that the interstitial electrons are genuinely populated near the Fermi level rather than being an artifact of a single integration window.

As illustrated in Fig.~\ref{fig:workflow}c, we subsequently retained the high-quality electrides by requiring a substantial interstitial volume fraction. For each PARCHG file, this fraction is defined as $f = 100\times V_\text{interstitial}/V_\text{cell}$ (\%), where $V_\text{interstitial}$ is the summed volume of the interstitial (non-nuclear) Bader basins and $V_\text{cell}$ is the unit cell volume. Specifically, we required $\max(e_{0.025}, e_{0.5}, e_{1.0}) \geq 20$\% and $\max(band_0, band_1) \geq 20$\%. This threshold is adopted from the volume-based topological analysis of Zhu~et al.~\cite{zhuComputationalDiscoveryInorganic2019} and ensures that the interstitial electrons occupy a significant fraction of the unit cell, comparable to the volume fraction of a conventional anionic site, rather than being a negligible charge-transfer artifact. We verified that confirmed electrides comfortably exceed this threshold in the identical pipeline.

Using the initial detection criterion above, we identified a total of 1,293 binary and 8,903 ternary electride candidates, requiring approximately 20,000 CPU hours.

\subsection*{Stage 4: Refined DFT Calculation}

For promising electride candidates identified from stage 3, we further perform structural relaxation until it reaches the full convergence. These relaxed structures were recomputed using the PBE functional according to Materials Project's setting to obtain more accurate energy values, which ensures the correctness of the $E_{\text{ref-hull-DFT}}$ as compared to the reference values in the Materials Project. At the stage, we also updated the reference convex hull with the new results, as illustrated in Fig. \ref{fig:workflow}d. This eventually leads to the identification of 232 binary and 32 ternary electride compounds with the $E_\text{hull-DFT}$ values less than 0.05~eV/atom. This stage costs approximately 38,000 CPU hrs. And all computed structural and property data are managed via the ASE's database utility \cite{ase}. We also computed phonons for a list of selected structures using the finite displacement method as implemented in the \texttt{Phonopy} code \cite{Togo-PRB-2008} to confirm the dynamical stability.

In general, this hierarchical screening framework strategically integrates physical principles with AI/ML tools for computational efficiency. First, we restrict our search to electron-rich compositions containing electropositive metals, thus greatly reducing combinatorial complexity. Second, \texttt{\texttt{MatterGen}} and \texttt{MatterSim} are used to rapidly explore and pre-screen candidates, achieving significant speedup over conventional DFT approaches. Finally, a two-stage DFT validation protocol further reduces computational costs by avoiding unnecessary calculation on non-interesting structures. This workflow efficiently identifies promising electride candidates across vast chemical space while maintaining rigor necessary for materials discovery.

\begin{figure*}[htbp]
\centering
\includegraphics[width=0.90\textwidth]{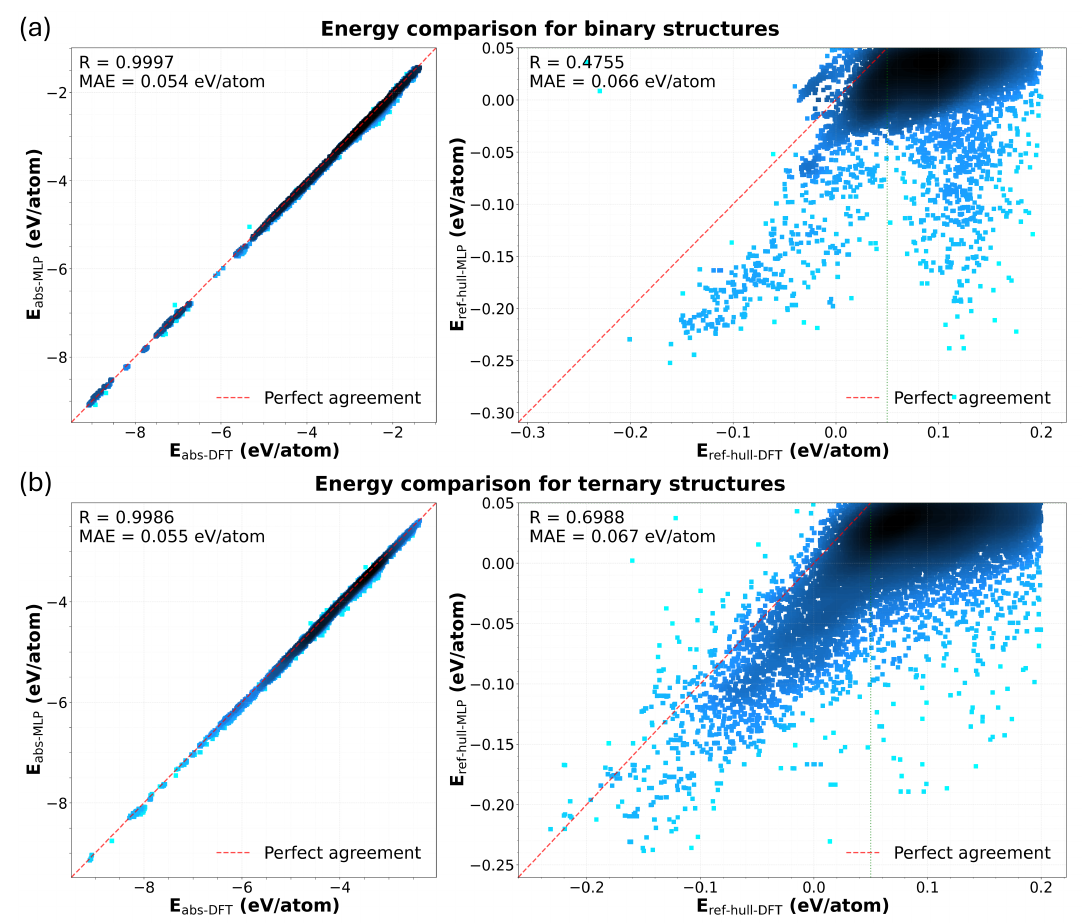}
\vspace{-2mm}
\caption{Validation of MLP Energy used in the screening workflow for (a) binary and (b) ternary systems, respectively. In both plots, the left panel shows the comparison in terms of absolute energy per atom ($E_\text{abs}$) between MLP and DFT results, while the right panel shows the comparison between energy with respect to the reference convex hull values ($E_\text{ref-hull}$) from the available materials project data.}
\label{fig:validation}
\vspace{-3mm}
\end{figure*}

\section{Results and Discussions}\label{sec:results}

\subsection{Validation of Hierarchical Screening Workflow}
As illustrated in Fig.~\ref{fig:workflow}, the workflow reduces the candidate pool from 79,244 to 17,195 for binary and from 285,120 to 18,705 for ternary systems after the MLP prescreen, and then 232 binary and 32 ternary candidates after DFT calculation. 
Such progressive pruning is expected, since electride formation is intrinsically rare. Hence, most generated electron-rich structures are either thermodynamically unfavorable or lack significant interstitial electron localization. In this workflow, the most critical assumption is that our MLP prescreen can effectively discard unstable structures while preserving the vast majority of genuinely promising candidates for the subsequent, more expensive DFT validation.

To verify the MLP accuracy, we compared both absolute energy $E_\text{abs}$ and $E_\text{ref-hull}$ values from \texttt{MatterSim}-MLP and PBE-DFT in Fig.~\ref{fig:validation} for 17,195 binary and 18,705 ternary structures generated at stages 2 and 3, respectively. 
Between \texttt{MatterSim}-MLP and VASP-DFT calculated absolute energy values for both binary and ternary compounds, Fig.~\ref{fig:validation} demonstrates the calculated metrics: mean absolute errors (MAE) 0.054/0.055~eV/atom, correlation coefficient R~=~0.9997/0.9986 for binary and ternary compounds are consistent with the report as found in the original literature \cite{yang2024mattersim}. The ternary system yields slightly bigger errors, which is expected due to the nature of MLP model construction.

Compared to $E_\text{abs}$, $E_\text{hull}$ is more important in our materials filtering strategy to determine whether a materials is energetically favorable. While many metastable structures can exist, it is generally considered that $E_\text{hull} < 0.10$~eV/atom is necessary for most compounds except nitrides \cite{sun2016thermodynamic}. Different from $E_\text{abs}$,  $E_\text{hull}$ is a relative quantity with respect to the reference systems, and thus it is more numerically sensitive. According to Fig. \ref{fig:validation}, the comparison of $E_\text{ref-hull}$ values yields similar MAE metrics, 0.066 eV/atom for binaries and 0.067 eV/atom for ternaries), while the correlation R value is much smaller (0.4755 for binary and 0.6988 for ternary). Although the correlation is not ideal, it is generally found that \texttt{MatterSim} tends to underestimate, rather than overestimate, the $E_\text{hull}$ values as compared to the DFT reference. Hence, the MLP prescreen may still consider a fraction of energetically unfavorable structures, but is less likely to miss the vast majority of truly low-energy candidates (E$_{\text{ref-hull-DFT}} < 0.05$~eV/atom), within the 0.10~eV/atom threshold value. Despite the underestimation issue, the use of MLP can effectively pre-relax the structure and filter approximately 60\% of unstable structures before more expensive VASP calculations, thus saving computational time in our screening framework.

\subsection{Binary Electride Discovery}

\begin{figure*}[htbp]
\centering
\includegraphics[width=0.85\textwidth]{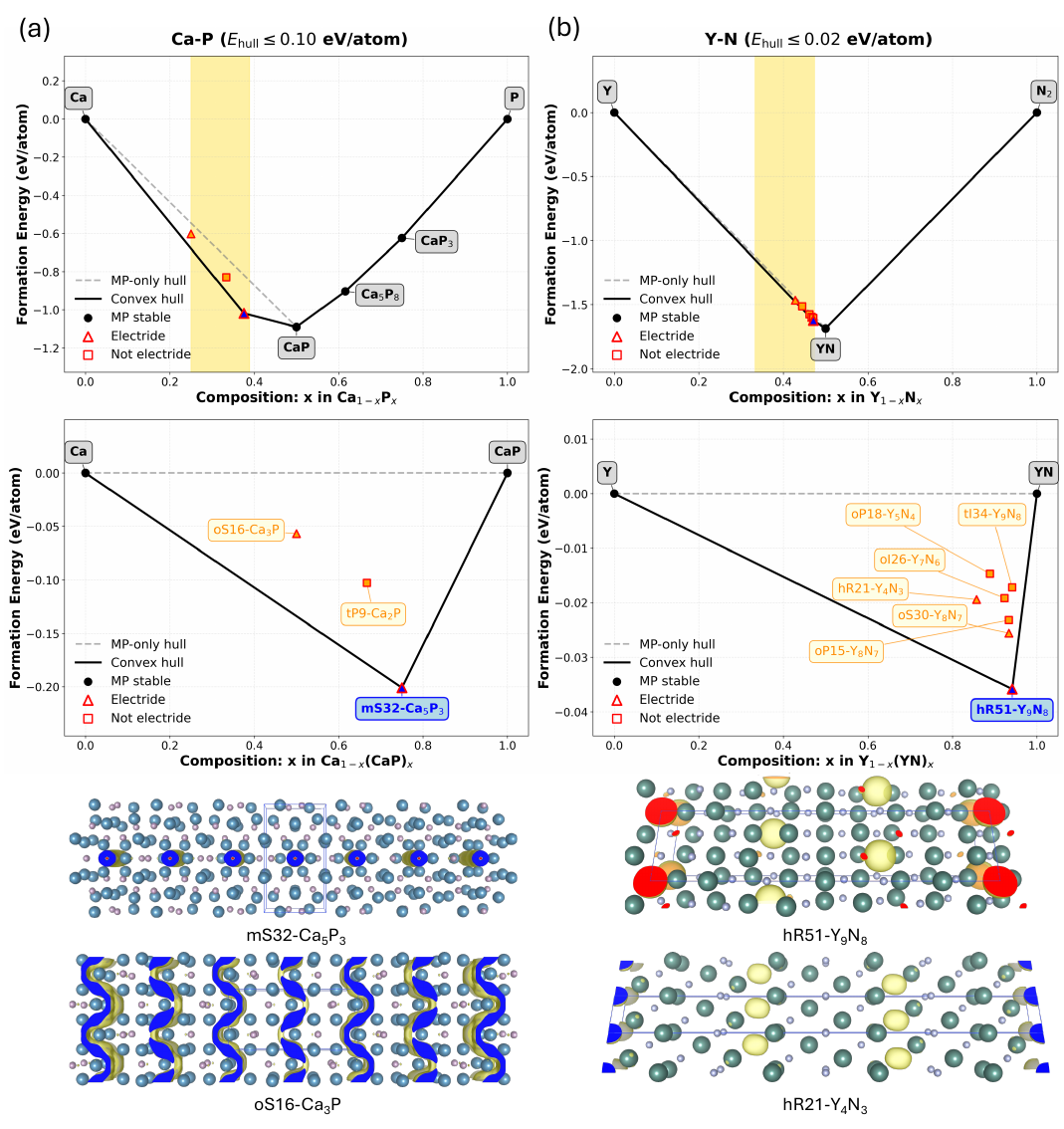}
\vspace{-3mm}
\caption{Convex hulls and new phases for (a) Ca-P and (b) Y-N systems. Top panels show the Materials Project (black circles) and newly discovered phases (red triangles for electrides and red squares for nonelectrides, on-hull structures are filled with blue, above hull structures are filled with orange), with the gray dashed line indicating the original convex hull and the solid black line representing the refined convex hull. The electron-rich compositions with $0 < N_\text{excess} \leq 4$ are highlighted in yellow rectangles. The middle panels display a zoomed-in sub-convex hull plot between the elemental metal and the stoichiometric binary with the most negative formation energy. The bottom panels illustrate representative structures from each system with atomic colors: Ca (steel blue), P (pale purple), Y (dark green), N (light blue); yellow isosurfaces represent excess electrons' distributions. The $E_\text{hull}$ criteria are used for DFT values, and per-panel energy values are display windows.}
\label{fig:binary_hulls}
\vspace{-3mm}
\end{figure*}

\begin{figure*}[htbp]
\centering
\includegraphics[width=0.85\textwidth]{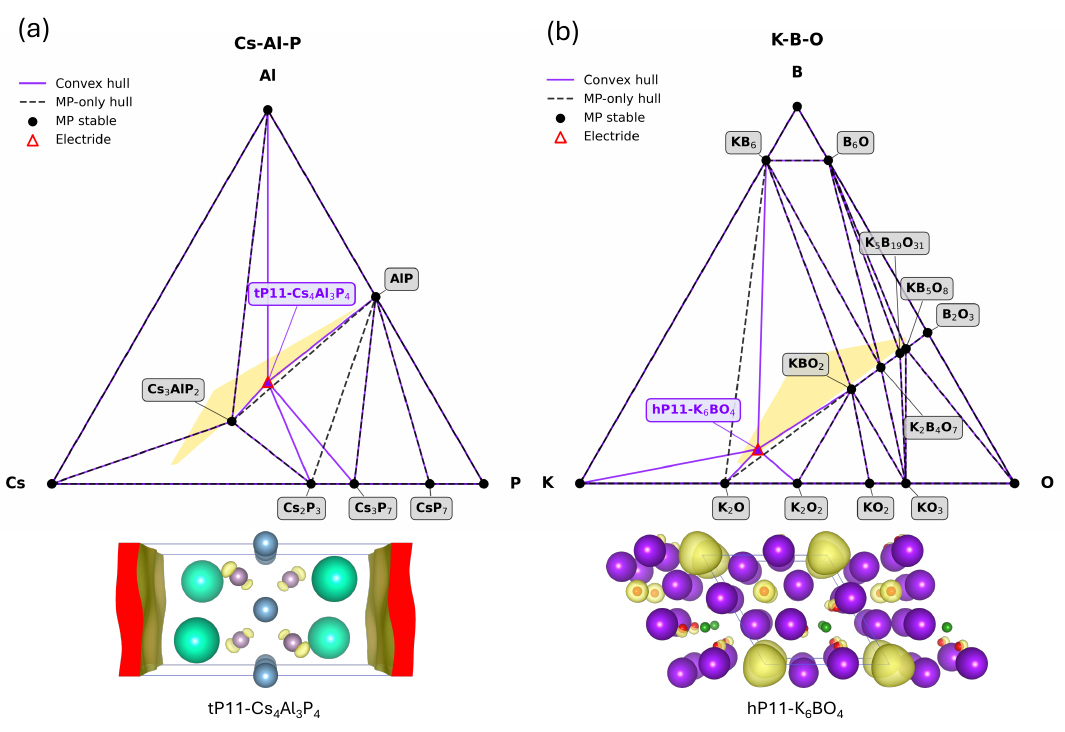}
\vspace{-3mm}
\caption{Ternary convex hulls and new phases for (a) Cs-Al-P and (b) K-B-O systems. Top panels: dashed black lines indicate original convex hull boundaries; solid purple lines show updated convex hulls after incorporating newly identified electride compounds. Black circles represent Materials Project phases; red triangles denote electride candidates (stable ones are filled with purple). The electron-rich composition region ($0 < N_{\text{excess}} \leq 2$) is highlighted by yellow shading. Bottom panels: representative electride structures from each system with atomic colors: K (purple), Cs (cyan), Al (gray), B (green), O (red), P (pale purple); yellow isosurfaces represent excess electrons' distributions. For these two specific ternary systems, there are no metastable candidates that fall within the $E_{\text{hull}}<0.05$~eV/atom discovery threshold. The complete set of generated electride candidates, including the metastable ones within the discovery threshold for all 15 ternary systems, is provided Figs.~S5--S6}
\label{fig:ternary_hulls}
\vspace{-3mm}
\end{figure*}

Among the screened 1,510 binary compositions, we have identified 232 low-energy candidates with E$_{\text{hull-DFT}} < 0.05$~eV/atom (see Figs. S1-S4~\cite{SM}). Below, we will briefly showcase the discoveries on two representative systems.

The Ca-P binary was reported to have three stable non-electride phases in the Materials Project, including Ca$_5$P$_8$ ($C2/m$: 12), CaP ($P\overline{6}2m$: 189) and CaP$_3$ ($P\overline{1}$: 2). As shown in Fig.~\ref{fig:binary_hulls}a, our search supplements 3 new candidates with $E_\text{ref-hull-DFT} < 0.10$~eV/atom. Remarkably, we identify one new stable phase that was absent from the MP database: mS32-Ca$_5$P$_3$ ($C2/m$: 12), lying 0.205~eV/atom below the original convex hull. This structure is closely  related to the well-known 1D-Mn$_5$Si$_3$ type~\cite{Lu-JACS-2016}, but crystallizes in a lower subgroup with the monoclinic symmetry. In addition, two metastable candidates were found to have $E_{\text{hull-DFT}} < 0.10$~eV/atom in the updated convex hull. In particular, the oS16-Ca$_3$P structure ($Cmma$: 65) is an electride with 0.077~eV/atom above the hull. Unlike Ca$_5$P$_3$, the oS16-Ca$_3$P forms a 2D corrugated interstitial electron distribution, suggesting a versatile electride chemistry in this system.

Additionally, we identified a dozen chemical systems exhibiting numerous low-energy structures clustered within electron-rich compositional regions. The Y-N system, shown in Fig.~\ref{fig:binary_hulls}b, exemplifies particularly rich phase behavior, with 7 candidates concentrated within a narrow energy range of $E_{\text{hull-DFT}} \leq 0.02$~eV/atom. Notably, one new thermodynamically stable phase, hR51-Y$_9$N$_8$ ($R\overline{3}$: 148), lies 0.036~eV/atom below the original Materials Project convex hull. This hR51-Y$_9$N$_8$ structure is characterized by nitrogen vacancies compared to the stoichiometric YN compound.
Between Y and Y$_9$N$_8$, six additional metastable candidates cluster within 0.02~eV/atom of the hull, including multiple Y$_8$N$_7$ and Y$_4$N$_3$ layered phases ($R\overline{3}m$: 166). These results collectively suggest that a thermodynamical route to achieve interstitial electron localization around Y by engineering the nitrogen vacancies in the existing YN compounds.

\subsection{Ternary Electride Discovery}

Extending from binary AB to ternary AA$^\prime$B compositions (A$^\prime$ =~Al, Si, Ge, Ga, Sn) identified 32 low-energy electride candidates with E$_{\text{hull-DFT}} < 0.05$~eV/atom (see details in Figs. S5 and S6~\cite{SM}). The small number is expected since our target compositions are slightly apart from the ideal stoichiometry. Hence, most of these compounds are unlikely to be energetically favorable unless the electride phases are formed. Below we highlight discoveries in two representative systems.

According to Materials Project, the Cs-Al-P system contains only one stable ternary phase: Cs$_3$AlP$_2$. Our results demonstrate that tP11-Cs$_4$Al$_3$P$_4$ ($P\bar42m$: 111), with one excess electron per formula unit, is also thermodynamically stable with respect to the existing compounds, with $E_\text{ref-hull-DFT}$ = -0.004~eV/atom). 
Fig.~\ref{fig:ternary_hulls}a shows the updated ternary phase diagram incorporating tP11-Cs$_4$Al$_3$P$_4$. From the diagram, tP11-Cs$_4$Al$_3$P$_4$ lies close to the compositional path between the two stoichiometric compounds Cs$_3$AlP$_2$ and AlP. Examination of the crystal structure reveals a layered packing with each layer terminated by the electron-positive Cs atoms on both sides. Consequently, it is expected that the excess electron localization between layers stabilizes this composition, similar to the behavior observed in 2D Ca$_2$N-type electrides. In addition to tP11-Cs$_4$Al$_3$P$_4$, we have found a total of 6 metastable (within 0.05~eV/atom $E_{\text{hull}}$) layered ternary electride candidates across a range of systems, suggesting a rich opportunity to achieving layered electrides in multi-composition compounds.

Our search does not only discover new materials in the underexplored systems, but also uncovers previously overlooked phases. The K-B-O system has been extensively studied due to its technological importance in lubricants and anti-corrosion applications, evidenced by the existence of 12 ternary entries in the Materials Project. However, our screening identified one new stable ternary electride: hP11-K$_6$BO$_4$ ($P3$: 143) with $E_{\text{ref-hull-DFT}}$ = -0.025~eV/atom (see Fig.~\ref{fig:ternary_hulls}b). This structure possesses the 1D cavities surrounded by potassium atoms, naturally providing a favorable environment for electron localization. This example demonstrates that generative approaches can identify overlooked compositions even in the extensively studied chemical systems.

\begin{figure*}[htbp]
\centering
\includegraphics[width=0.9\textwidth]{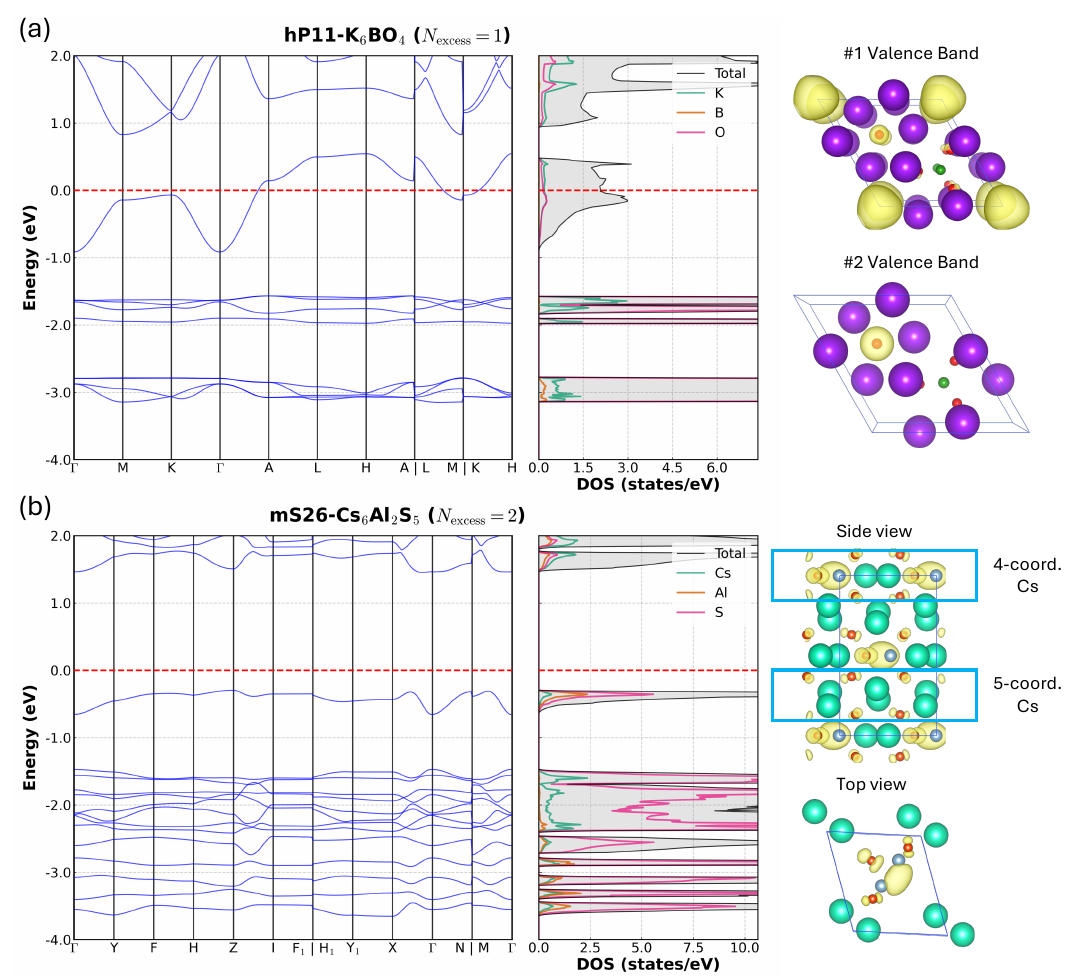}
\vspace{-3mm}
\caption{Electronic structures of two representative stable ternary electrides: (a) hP11-K$_6$BO$_4$ and (b) mS26-Cs$_6$Al$_2$S$_5$. 
Left panels: band structures and projected density of states (PDOS). Right panels: partial charge density distributions for the top valence bands in hP11-K$_6$BO$_4$ and mS26-Cs$_6$Al$_2$S$_5$ (shown in top and side views). Atomic colors: K (purple), Cs (cyan), Al (gray), B (green), O (red), S (orange); yellow isosurfaces represent the electron charge density.}
\label{fig:band_dos_plot}
\vspace{-3mm}
\end{figure*}

\subsection{Electronic Structure Analysis}
Among the 264 identified low-energy electrides, we have found 13 thermodynamically stable electrides, 12 of which exhibit non-imaginary phonon frequencies (see Figs. S7-S11), suggesting dynamical stability and favorable potential synthetic accessibility. In Fig.~\ref{fig:band_dos_plot}, we examine the electronic structures of two representative stable ternary electrides, hP11-K$_6$BO$_4$ ($P3$: 143) and mS26-Cs$_6$Al$_2$S$_5$ ($Cm$: 8), which feature different excess electron counts.

Fig.~\ref{fig:band_dos_plot}a presents the electronic band structure and atom-projected density of states (PDOS) for hP11-K$_6$BO$_4$. Given one excess electron per primitive unit cell exist in this compound, we expect the excess electron to partially occupy the highest valence band if electride formation occurs. Indeed, the band structure, show in the left panel of Fig.~\ref{fig:band_dos_plot}a, exhibits a characteristic half-filled top valence band. The PDOS analysis (middle panel) reveals that this valence band has negligible projections onto K, B, or O atomic orbitals, demonstrating typical interstitial electron character. The computed partial charge density for the top valence band (right panel) displays a spherical distribution centered at the hexagonal lattice origin (0, 0, 0), consistent with nearly free electron behavior. This interpretation is further supported by the parabolic band dispersion near the $\Gamma$ point, characteristic of a quasi-free electron gas similar to previously identified switchable 2D electrides~\cite{Yang-PRB-2021}. In contrast, the second-highest valence band is primarily composed of oxygen's $p$-orbitals, positioned approximately 0.75 eV below the interstitial band, indicating clear electronic separation between interstitial and atomic-like states.

If one excess electron leads to a half-filled interstitial band, one may naturally speculate that two excess electrons result in a fully filled interstitial band. Indeed, the electronic structure of mS26-Cs$_6$Al$_2$S$_5$ in Fig.~\ref{fig:band_dos_plot}b supports this hypothesis. With two excess electrons per primitive unit cell, mS26-Cs$_6$Al$_2$S$_5$ displays a fully filled top valence band, resulting in a semiconducting electride with a PBE band gap of $\sim$1.75 eV. However, the PDOS analysis reveals a more nuanced scenario than in hP11-K$_6$BO$_4$. The top valence band contains not only interstitial electrons but also notable contributions from sulfur's $p$-orbitals. The partial charge density plot of the top valence band (right panel of Fig.~\ref{fig:band_dos_plot}b) agrees with this observation, showing electron density distributed between the hexagonal interstitial center and around S atoms. This complex electronic structure arises from the crystal packing, where the structure consists of alternating electron-neutral layers (with 5-coordinated Cs) and electron-rich layers (with 4-coordinated Cs). Within the electron-rich layers, the under-coordinated Cs atoms form a nearly hexagonal arrangement in-plane, with each Cs donating excess electrons. However, the presence of S and Al atoms causes some electrons to occupy sulfur $p$-orbitals, with the majority localizing at the hexagonal interstitial centers. It is worth noting the interstitial space in mS26-Cs$_6$Al$_2$S$_5$ is more spatially confined due to the existence of in-plane Al and S atoms. If a more spacious interstitial environment were available, one could potentially achieve a fully interstitial-character valence band without PDOS on the atomic sites. That said, mS26-Cs$_6$Al$_2$S$_5$ still exhibits clear electride character, as confirmed by the significant interstitial electron localization at the top valence band. Similar to hP11-K$_6$BO$_4$, the second highest valence band is composed on sulfur's $p$-orbitals with about 1.0 eV separation from the top valence band. This plot is not shown here to maintain the clarity.

In our screening, most electride compounds exhibit $N_\text{excess}=1$, corresponding to a half-filled top valence band, consistent with the majority of previously identified electrides. This observation suggests two complementary avenues for future exploration. First, restricting the target composition space to lower $N_\text{excess}$ values could enable more efficient systematic exploration of extended chemical spaces beyond ternary compositions. Second, screening compositions with $N_\text{excess}=2$ offers greater likelihood for identifying intrinsically semiconducting electrides with enhanced thermodynamic stability for applications in semiconducting devices.. These insights point toward more targeted screening strategies for expanding the electride materials landscape.

\section{Conclusion}\label{sec:conclusion}
In this work, we have presented an accelerated framework for inorganic electride discovery by integrating physical principles, generative AI, ML potentials, and high-throughput DFT validation. Compared to the traditional framework based on DFT calculations, our approach can efficiently explore vast chemical space with significantly reduced computational costs, while maintaining the necessary accuracy. Using this workflow to a targeted chemical space of electron-rich binary and ternary compositions, we have identified 264 new electride candidates within 0.05~eV/atom above the convex hull, substantially expanding the known electride landscape. 

We highlight the following key components in our work.

First, chemical space selection based on physical principles is essential for efficient materials discovery. Rather than uniformly exploring all possible compositions, restricting the search to electron-rich systems with electropositive metals dramatically reduces combinatorial complexity while maintaining access to chemically reasonable candidates without sacrificing discovery potential.

Second, the use of \texttt{MatterGen} and \texttt{MatterSim} allows rapid generation and screening of low-energy structures across thousands of compositions. This approach achieves a significant speed up to enable systematic exploration of chemical space previously inaccessible to computational materials discovery.

Third, the identification of new stable phases demonstrates that generative methods can uncover compositional variants and structural prototypes overlooked by previous database-mining approaches, suggesting substantial opportunities for targeted re-examination of existing materials spaces.

Finally, these discoveries underscore the transformative potential of modern AI and ML tools in materials discovery. However, it is critical to acknowledge that both \texttt{MatterGen} and \texttt{MatterSim} were trained on existing data, which may introduce biases toward chemically conventional compositions and known structural families. Future work should explore training generative models on more diverse datasets.

Our framework combining physical insights, generative modeling, and ML prescreening provides a roadmap for accelerated discovery across diverse functional materials beyond electrides, establishing a new paradigm for materials exploration in the age of AI-driven science.
Transferring the workflow to a new target requires adapting mainly two components: (1) a composition-level physical descriptor that constrains the chemical space (analogous to the excess-electron count used here), and (2) a property analyzer applied in the final high-throughput DFT stage; the generative sampling, ML-potential prescreening, and hierarchical DFT validation are reused unchanged. With these minimal changes, the same pipeline can be applied to, for example, thermoelectrics (screening for low lattice thermal conductivity or favorable power factor)~\cite{ghafarollahi2025autonomous, guo2025generative, longInverseDesignHighperformance2025}, conventional superconductors (electron--phonon coupling, to which the free-electron-like states of electrides are themselves germane)~\cite{zhangElectrideSuperconductivityBehaviors2017, zhangSuperconductivityElectride2023}, topological semimetals and insulators (band-inversion descriptors, again naturally connected to weakly bound interstitial electrons)~\cite{hirayamaElectridesNewPlatform2018, Park-PRL-2018, huangTopologicalElectrideY2C2018, nieApplicationTopologicalQuantum2021}, low-work-function electron emitters~\cite{drobny2024endurance}, catalysts~\cite{Kitano-NChem-2012, Kitano-NC-2015}, and magnetic materials~\cite{zeni2025generative}.
The identified candidates may stimulate further experimental efforts to assess their practical utility for applications.

\section*{Acknowledgments}
This research was sponsored by the U.S. Department of Energy, Office of Science, Office of Basic Energy Sciences, and the Established Program to Stimulate Competitive Research (EPSCoR) under the DOE Early Career Award No. DE-SC0024866, as well as the UNC Charlotte - Division of Research Postdoctoral Hiring Program.

\section*{Data availability}
The source code, instructions, as well as scripts used to calculate the results of this study, are available in \url{https://github.com/MaterSim/ElectrideFlow}. The newly identified electrides are also interactively available via \url{https://mmi.charlotte.edu/electride}.

\section*{Conflict of interest}
All authors declare that they have no conflict of interest.

\bibliographystyle{apsrev4-2}
\bibliography{main}

\end{document}